\documentclass[slac_one]{revtex4}
\usepackage{graphicx}
\usepackage{fancyhdr}
\begin{document}

\title{The Constrained E$_6$SSM} 

%

\author{P. Athron}
\affiliation{Institut f\"ur Kern- und Teilchenphysik, TU Dresden, Dresden, D-01062, Germany}
\author{S. F. King, S. Moretti}
\affiliation{University of Southampton, Southampton, SO17 1BJ, UK}
\author{D. J. Miller, R. Nevzorov}
\affiliation{University of Glasgow, Glasgow, G12 8QQ, UK}

\begin{abstract}
We present a self--consistent $E_6$ inspired supersymmetric model with an extra 
$U(1)_{N}$ gauge symmetry under which right-handed neutrinos have zero charge.
We explore the particle spectrum within the constrained version of this 
exceptional supersymmetric standard model (E$_6$SSM) and discuss its possible
collider signatures.
\end{abstract}

\maketitle

\thispagestyle{fancy}


\section{INTRODUCTION}
The unification of gauge couplings in the supersymmetric (SUSY) models
allows one to embed the gauge group of the standard model (SM) into Grand 
Unified Theories (GUTs) based on simple gauge groups such as $SU(5)$, 
$SO(10)$ or $E_6$. On the other hand gauge coupling unification 
permits one to incorporate SUSY models into superstring theories that make 
possible partial unification of gauge interactions with gravity. 
At high energies the $E_6$ symmetry in the superstring inspired models 
can be broken to rank--5 subgroup $SU(3)_C\times SU(2)_W\times U(1)_Y\times U(1)'$ 
where $U(1)'=U(1)_{\chi} \cos\theta+U(1)_{\psi} \sin\theta$.
Two anomaly-free $U(1)_{\psi}$ and $U(1)_{\chi}$ symmetries originate from
the breakdown of $E_6$ and $SO(10)$ respectively, i.e. $E_6\to SO(10)\times U(1)_{\psi},~
SO(10)\to SU(5)\times U(1)_{\chi}$. Here we concentrate on a particular $E_6$ 
inspired supersymmetric model with an extra $U(1)_{N}$ gauge symmetry that corresponds $\theta=\arctan\sqrt{15}$. Only in this exceptional supersymmetric standard model (E$_6$SSM) 
\cite{e6ssm} the right--handed neutrinos do not participate in gauge interactions. 
Therefore they may be superheavy shedding light on the origin of lepton mass 
hierarchy. The extra $U(1)_{N}$ gauge symmetry survives to low energies and
forbids a bilinear term $\mu \hat{H}_d\hat{H}_u$ in the superpotential of the 
considered model but allows the interaction $\lambda S H_d H_u$. At the electroweak (EW)
scale the scalar component of the SM singlet superfield $S$ acquires a non-zero 
vacuum expectation value (VEV), $\langle S \rangle=s/\sqrt{2}$, breaking $U(1)_N$ 
and an effective $\mu=\lambda s/\sqrt{2}$ term is automatically generated. 
Thus the $\mu$ problem in the E$_6$SSM is solved in a similar way to the next-to-minimal 
supersymmetric standard model (NMSSM), but without the accompanying problems of singlet 
tadpoles or domain walls.

\section{THE E$_6$SSM}
The E$_6$SSM is based on the $SU(3)_C\times SU(2)_W\times U(1)_Y \times U(1)_N$ 
gauge group which is a subgroup of $E_6$. To ensure anomaly cancellation the particle
content of the E$_6$SSM is extended to include three complete {\bf $27$} representations 
of $E_6$ \cite{e6ssm}. Each {\bf $27_i$} multiplet contains a SM family of quarks and leptons, 
right--handed neutrino $N^c_i$, SM singlet field $S_i$ which carries a non--zero $U(1)_{N}$ 
charge, a pair of $SU(2)_W$--doublets $H_{1i}$ and $H_{2i}$ with the quantum numbers 
of Higgs doublets and a pair of colour triplets of exotic quarks $\overline{D}_i$ and $D_i$
which can be either diquarks (Model I) or leptoquarks (Model II). $H_{1i}$ and $H_{2i}$ 
form either Higgs or inert Higgs multiplets. We also require a further pair $H'$ and 
$\overline{H}'$ from incomplete extra $27'$ and $\overline{27'}$ representations to survive 
to low energies to ensure gauge coupling unification. Our analysis reveals that 
the unification of the gauge couplings in the E$_6$SSM can be achieved for any phenomenologically 
acceptable value of $\alpha_3(M_Z)$, consistent with the central measured low energy value, 
unlike in the MSSM which requires significantly higher values of $\alpha_3(M_Z)$, well above central
measured value \cite{unif-e6ssm}.

Since right--handed neutrinos have zero charges in the considered model they can acquire 
very heavy Majorana masses. The heavy Majorana right-handed neutrinos may decay into final states 
with lepton number $L=\pm 1$, thereby creating a lepton asymmetry in the early Universe. 
Because the Yukawa couplings of exotic particles are not constrained by the neutrino oscillation 
data the substantial values of the CP asymmetries in the considered model can be induced even 
for a relatively small mass of the lightest right--handed neutrino ($M_1 \sim 10^6\,\mbox{GeV}$) 
so that the successful thermal leptogenesis may be achieved without encountering gravitino 
problem \cite{leptogen-e6ssm}. 

The superpotential of the E$_6$SSM involves a lot of new Yukawa couplings in comparison to the SM. 
In general these new interactions violate baryon number conservation and induce non-diagonal 
flavour transitions. To suppress baryon number violating and flavour changing processes one can 
postulate a $Z^{H}_2$ symmetry under which all superfields except one pair of $H_{1i}$ and $H_{2i}$ 
(say $H_d\equiv H_{13}$ and $H_u\equiv H_{23}$) and one SM-type singlet field ($S\equiv S_3$) are odd. 
The $Z^{H}_2$ symmetry reduces the structure of the Yukawa interactions to:
\begin{equation}\label{2}
\begin{array}{c}
W_{\rm E_6SSM}\longrightarrow \lambda_i S(H_{1i}H_{2i})+
\kappa_i S(D_i\overline{D}_i)+f_{\alpha\beta}S_{\alpha}(H_d H_{2\beta})+
\tilde{f}_{\alpha\beta}S_{\alpha}(H_{1\beta}H_u)\\[0mm]
+h^{E}_{4j}(H_d H')e^c_j
+\mu'(H'\overline{H}')+W_{\rm{MSSM}}(\mu=0),
\end{array}
\end{equation}
where $\alpha,\beta=1,2$ and $i=1,2,3$\,. Here we assume that all right--handed neutrinos are relatively 
heavy so that they can be integrated out. The $SU(2)_W$ doublets $H_u$ and $H_d$, that are even 
under $Z^{H}_2$ symmetry, play the role of Higgs fields generating the masses of quarks and leptons after 
the EW symmetry breaking (EWSB). The singlet field $S$ must also acquire large VEV to induce sufficiently 
large masses for the exotic charged fermions and $Z'$ boson to avoid conflict with direct particle searches 
at present and former accelerators. This implies that the Yukawa couplings $\lambda_i$ and $\kappa_i$ 
should be large enough so that the evolution of the soft scalar mass $m_S^2$ of the singlet 
field $S$ results in negative values of $m_S^2$ at low energies, triggering the 
breakdown of $U(1)_{N}$ symmetry. To guarantee that only $H_u$, $H_d$ and $S$ acquire VEVs in the E$_6$SSM 
a certain hierarchy between the Yukawa couplings is imposed, i.e. $\kappa_i\sim\lambda_i\gg 
f_{\alpha\beta},\tilde{f}_{\alpha\beta},h^{E}_{4j}$.

After the breakdown of the gauge symmetry $H_u$, $H_d$ and $S$ form three CP-even, one CP-odd and two charged 
states in the Higgs spectrum. 
The mass of the one CP--even Higgs particle is always very close to the $Z'$ 
boson mass $M_{Z'}$. The masses of another CP--even, CP--odd and charged Higgs states are almost degenerate. 
As in the MSSM and NMSSM one of the CP--even Higgs bosons is always light irrespective of the SUSY breaking scale. 
However in contrast with the simplest SUSY models the lightest Higgs boson in the E$_6$SSM can be heavier than 
$110-120\,\mbox{GeV}$ even at the tree level. In the two--loop approximation the lightest Higgs boson mass does not 
exceed $150-155\,\mbox{GeV}$ \cite{e6ssm}. Thus the SM--like Higgs boson in the E$_6$SSM can be considerably heavier 
than in the MSSM and NMSSM. 

\begin{table}[t]
\begin{center}
\caption{Particle spectrum in the constrained E$_6$SSM for $\tan\beta=10$
(All mass parameters are given in GeV).}
\begin{tabular}{|c|c|c|}
\hline 
                         & \textbf{Scenario I} & \textbf{Scenario II} \\
\hline 
$\lambda_3(M_X)$         &   -2.0              &   -0.395\\
$\lambda_{1,2}(M_X)$     &    2.6              &    0.1\\
$\kappa_3(M_X)$          &    2.5              &    0.43\\
$\kappa_{1,2}(M_X)$      &    2.5              &    0.08\\
$s$                      &    4000             &    2700\\
$M_{1/2}$                &    389              &    358 \\
$m_0$                    &    725              &    623 \\
$A$                      &    -1528            &    757 \\
\hline 
$m_{\tilde{D}_{1}}(3)$   &    1948             &    1445\\
$m_{\tilde{D}_{2}}(3)$   &    2200             &    2059\\
$\mu_D(3)$               &    2060             &    1747\\
$m_{\tilde{D}_{1}}(1,2)$ &    1948             &    370 \\
$m_{\tilde{D}_{2}}(1,2)$ &    2200             &    916 \\
$\mu_D(1,2)$             &    2060             &    391 \\
\hline
$|m_{\chi^0_6}|$         &    1548             &    1051\\
$m_{h_3}\simeq M_{Z'}$   &    1518             &    1021\\
$|m_{\chi^0_5}|$         &    1490             &    994 \\
\hline
$m_S(1,2)$               &    1290             &    961 \\
$m_{H_2}(1,2)$           &    1172             &    561 \\
$m_{H_1}(1,2)$           &    903              &    345 \\
$\mu_{\tilde{H}}(1,2)$   &    1302             &    229 \\
\hline
$m_{\tilde{u}}(1,2)$     &    1007             &    845 \\
$m_{\tilde{d}}(1,2)$     &    1113             &    903 \\
$m_{\tilde{Q}}(1,2)$     &    1023             &    862 \\
$m_{\tilde{L}}(1,2,3)$   &    1015             &    796 \\
$m_{\tilde{e}}(1,2,3)$   &     873             &    708 \\
$m_{\tilde{b}_2}$        &    1108             &    894 \\
$m_{\tilde{b}_1}$        &     907             &    712 \\
$m_{\tilde{t}_2}$        &     921             &    772 \\
$m_{\tilde{t}_1}$        &     777             &    474 \\
\hline
$|m_{\chi^0_3}|\simeq |m_{\chi^0_4}|\simeq |m_{\chi^{\pm}_2}|$ 
                         &     739             &    685 \\
$m_{h_2}\simeq m_A \simeq m_{H^{\pm}}$
                         &     615             &    720 \\
$m_{h_1}$                &     116             &    114 \\
\hline
$M_{\tilde{g}}$          &     350             &    327 \\
$|m_{\chi^{\pm}_1}|\simeq |m_{\chi^0_2}|$
                         &     106             &    101 \\
$|m_{\chi^0_1}|$         &     59              &    57  \\
\hline
\end{tabular}
\label{tabular}
\end{center}
\end{table}

\section{PARTICLE SPECTRUM IN THE CONSTRAINED E$_6$SSM}
The E$_6$SSM contains many new parameters. Even if we neglect $f_{\alpha\beta}$, $\tilde{f}_{\alpha\beta}$ and $h^{E}_{4j}$ 
the simplified superpotential of the E$_6$SSM involves seven extra couplings ($\mu'$, $\kappa_i$ and $\lambda_i$) as 
compared with the MSSM with $\mu=0$. The soft breakdown of supersymmetry gives rise to many new parameters. 
The number of fundamental parameters can be reduced drastically within the constrained version of the E$_6$SSM (cESSM).
Constrained SUSY models imply that all soft scalar masses are set to be equal to $m_0^2$ at the scale $M_X$, all gaugino 
masses $M_i(M_X)$ are equal to $M_{1/2}$ and trilinear scalar couplings $A_i(M_X)=A$. Thus cESSM is characterised by the set 
of Yukawa couplings, which are allowed to be of the order of unity, and universal soft SUSY breaking terms, i.e.
\begin{equation}\label{3}
\lambda_i(M_X),\quad \kappa_i(M_X),\quad h_t(M_X),\quad h_b(M_X), \quad h_{\tau}(M_X), \quad m_0, \quad M_{1/2},\quad A, 
\end{equation}
where $h_t(M_X)$, $h_b(M_X)$ and $h_{\tau}(M_X)$ are the $t$--quark, $b$--quark and $\tau$--lepton Yukawa couplings. 
To simplify our analysis we assume that all parameters (\ref{3}) are real and $M_{1/2}$ is positive. 
In order to guarantee the correct EWSB $m_0^2$ has to be positive. The set of the cESSM parameters (\ref{3}) should be 
in principle supplemented by $\mu'$ and the associated bilinear scalar coupling $B'$. However since
$\mu'$ is not constrained by the EWSB and the term $\mu'H'\overline{H}'$ in the superpotential (\ref{2})
is not suppressed by the $E_6$ the parameter $\mu'$ is expected to be $\sim 10\,\mbox{TeV}$ so that $H'$ and $\overline{H}'$ 
decouple from the rest of the particle spectrum. As a consequence parameters $B'$ and $\mu'$ are irrelevant for our analysis.

In order to calculate the particle spectrum within the cESSM we evolve all 
mass parameters from the Grand Unification scale to the SUSY breaking scale 
for each set of gauge and Yukawa couplings at the scale $M_X$. In our
analysis we use two--loop renormalisation group (RG) equations for the gauge and
Yukawa couplings together with two--loop RG equations for $M_a(\mu)$ and 
$A_i(\mu)$ as well as one--loop RG equations for $m_i^2(\mu)$.
At the next stage of our analysis we fix $s$ and $\tan\beta=v_2/v_1$, where $v_2$ and $v_1$
are the VEVs of the Higgs fields $H_u$ and $H_d$, and choose $A$, $m_0$ and $M_{1/2}$
so that the EW symmetry breaking constraints are satisfied, i.e.
\begin{equation}\label{5}
\frac{\partial V}{\partial s}=\frac{\partial V}{\partial v_1}=\frac{\partial V}{\partial v_2}=0\,,
\end{equation}
where $V$ is a Higgs effective potential. Although the correct
EWSB is not guaranteed in the considered model, remarkably there is always a solution for
sufficiently large values of $\kappa_i$ which drive $m_S^2$ negative. Finally at the last
stage of our analysis we vary Yukawa couplings, $\tan\beta$ and $s$ to establish the allowed
range of the parameters and qualitative pattern of the particle spectrum in the E$_6$SSM. 

The results of our study of the particle spectrum in the E$_6$SSM are summarised in the Table~1.
We find that a set of the lightest SUSY particles in the cE$_6$SSM always includes the gluino $\tilde{g}$, the 
two lightest neutralino ($\chi_1^0$ and $\chi_2^0$) and the lightest chargino $\chi^{\pm}_1$. 
All other SUSY particles can be substantially heavier. Nevertheless there exists a part of the 
E$_6$SSM parameter space where exotic quarks, exotic squarks, inert Higgs bosons and inert Higgsinos 
can be also relatively light (see Scenario II).

\section{COLLIDER SIGNATURES}
If neutralino, chargino and gluino are the only lightest SUSY particles then one can expect to observe 
$\tilde{g}\tilde{g}$, $\chi^{\pm}_1\chi^{\mp}_1$, $\chi_2^0 \chi^{\pm}_1$ and $\chi_2^0 \chi_2^0$
pair production at the LHC. Since each gluino will decay further into quark, antiquark and lightest 
neutralino (or chargino) the gluino pair production will result in an appreciable enhancement of the 
cross section of $pp\to q\bar{q}q\bar{q}+E^{miss}_{T}+X$\,. If all squarks are much heavier than the gluino,
i.e. particle spectrum has a very hierarchical structure, then gluinos can be relatively long lived 
particles because $\Gamma_{\tilde{g}}\propto M_{\tilde{g}}^5/m_{\tilde{q}}^4$ in the considered case.
In particular their lifetime can be comparable with the lifetime of $W^{\pm}$ and $Z$ bosons. At the same 
time the pair production of the second lightest neutralino also gives rise to a remarkable signature. 
When $\chi_1^0$, $\chi_2^0$ and $\chi^{\pm}_1$ are the lightest SUSY particles the second lightest 
neutralino will decay predominantly into lepton, antilepton and $\chi_1^0$. Thus $\chi_2^0 \chi_2^0$
pair production would generate an excess in the cross section of 
$pp\to l\overline{l} l\overline{l}+E^{miss}_{T}+X$ that should be possible to observe at the LHC.

Other possible manifestations of the E$_6$SSM at the LHC are related with the presence of a $Z'$ 
and of exotic multiplets of matter. For instance, $Z'$ boson can be discovered at the LHC
if its mass is less than $4-4.5\,\mbox{TeV}$. When the Yukawa couplings of the exotic particles 
have a hierarchical structure some of the exotic fermions can be relatively light so that 
their production cross section at the LHC can be comparable with the cross section of $t\bar{t}$ 
production. Assuming that $D_i$ and $\overline{D}_i$ couple most strongly with the third family 
quarks and leptons the lightest exotic quarks decay into either two heavy quarks $Q Q$ or 
a heavy quark and $\tau$--lepton $Q\tau (\nu_{\tau})$, where $Q$ is either a $b$- or 
$t$-quark. This can lead to the substantial enhancement of the cross section of either 
$pp\to Q\bar{Q}Q^{'}\bar{Q}^{'}+E^{miss}_{T}+X$ or $pp\to Q\bar{Q}l\bar{l}+E^{miss}_{T}+X$. 
The discovery of the $Z'$ and exotic quarks predicted by the E$_6$SSM would represent a possible 
indirect signature of an underlying $E_6$ gauge structure at high energies and provide a window 
into string theory.

\begin{acknowledgments}
RN acknowledges support from the SHEFC grant HR03020 SUPA 36878.
\end{acknowledgments}


\begin{thebibliography}{9}   

\bibitem{e6ssm}
S.F. King, S. Moretti, R. Nevzorov,
``Theory and phenomenology of an exceptional supersymmetric standard
model'', Phys.\ Rev.\  D {\bf 73} (2006) 035009;
S.F. King, S. Moretti, R. Nevzorov,
``Exceptional supersymmetric standard model'',
Phys.\ Lett.\  B {\bf 634} (2006) 278.

\bibitem{unif-e6ssm}
S.F. King, S. Moretti, R. Nevzorov,
``Gauge Coupling Unification in the Exceptional Supersymmetric Standard Model,''
Phys.\ Lett.\  B {\bf 650} (2007) 57.

\bibitem{leptogen-e6ssm}
S.F. King, R. Luo, D.J. Miller, R. Nevzorov,
``Leptogenesis in the Exceptional Supersymmetric Standard Model: flavour
dependent lepton asymmetries'', arXiv:0806.0330 [hep-ph].

\end{thebibliography}
\end{document}